\documentclass[a4paper,11pt]{article}
\usepackage{latexsym}
\usepackage{amsbsy}
\title{Quantum states of a generalized time-dependent \\ inverted harmonic oscillator}
\author{I. A. Pedrosa\footnote{Permanent address: Universidade Federal da Para\'{\i}ba, Departamento de F\'{\i}sica - CCEN, Caixa Postal 5008, CEP. 58051-970, Jo\~ao Pessoa, PB, Brazil, e-mail: iapedrosa@fisica.ufpb.br} { and} I. Guedes\footnote{e-mail:
guedes@fisica.ufc.br} \\Departamento de F\'{\i}sica, Universidade
Federal do Cear\'a, \\ Caixa Postal 6030, 60451-970, Fortaleza,
Cear\'a, Brazil}
\date{}
\begin{document}
\maketitle
\begin{abstract}
We discuss the extension of the Lewis and Riesenfeld method of
solving the time-dependent Schr\"odinger equation to cases where
the invariant has continuous eigenvalues and apply it to the case
of a generalized time-dependent inverted harmonic oscillator. As a
special case, we consider a generalized inverted oscillator with
constant frequency and exponentially increasing mass. \\ \\ PACS
number(s): 03.65.Ca, 03.65.Fd, 03.65.Ge.
\end{abstract}
$$\\
\textbf{I. INTRODUCTION}
$$
\par
In the last few years, the study of the time-dependent inverted
harmonic oscillator has attracted some attention in the
literature[1-4]\nocite{r1}\nocite{r2}. Despite its mathematical
interest, this problem may find interesting applications in
different areas of physics once many quantum-mechanical effects
can be treated phenomenologically by means of the time-dependent
parameters in the Hamiltonian of the system.

The time-dependent inverted harmonic oscillator is exactly
solvable just like the standard time-dependent harmonic
oscillator. However, the physics of the time-dependent inverted
oscillator is very different: it has a wholly continuous energy
spectrum varying from minus to plus infinity; its energy
eigenstates are no longer square-integrable and they are doubly
degenerate with respect to either the incident direction or,
alternatively, the parity.

The time-dependent inverted harmonic oscillator with an
exponentially increasing mass and constant frequency, the
so-called inverted Caldirola-Kanai oscillator, has been discussed
by Baskoutas et al\cite{r3}. On the other hand, the most general
case where mass and frequency are both time-dependent has been
considered by us in Ref.\cite{r4}.

In this paper, we discuss the Lewis and Riesenfeld (LR) invariant
method\cite{r5} to cases where the invariant has continuous
eigenvalues and apply it to obtain the exact wave functions of a
generalized time-dependent inverted harmonic oscillator. As a
special case, we also find the wave functions of a generalized
inverted oscillator with constant frequency and exponentially
increasing mass. In Sec. II, we briefly outline the LR invariant
method to cases where the invariant has continuous eigenvalues. In
Sec III, we find the Schr\"odinger wave functions for a
generalized time-dependent inverted oscillator. In addition, as a
particular case, we obtain the wave functions of a generalized
inverted oscillator with constant frequency and exponentially
increasing mass. We end with a summary in Sec. IV.
$$\\
\textbf{II. THE LR INVARIANT METHOD: CONTINUOUS EIGENVALUES}
$$
\par
The work of LR\cite{r5} assumes that the eigenvalue spectrum for
the invariant is discrete. Here, we wish investigate the LR method
in the case of continuous eigenvalues. To do so, we will consider
a quantum system characterized by a time-dependent Hamiltonian
$H(t)$ and whose corresponding Schr\"odinger equation is
\begin{equation}
\label{eq:n1}
i \hbar \frac{\partial \psi (q, \, t)}{\partial t} =
H(t) \psi (q, \, t).
\end{equation}
A Hermitian operator $I(t)$ is called an invariant for the system if it satisfies
\begin{equation}
\label{eq:n2}
\frac{\textrm{d}I}{\textrm{d}t} = \frac{1}{i \hbar}
[I, \, H] + \frac{\partial I}{\partial t} = 0.
\end{equation}
The eigenvalue equation for $I(t)$ can be written as\cite{r6}
\begin{equation}
\label{eq:n3}
I \phi_{\lambda} (q, \, t) = \lambda \phi_{\lambda}
(q, \, t),
\end{equation}
where
\begin{equation}
\label{eq:n4}
\left <\phi_{\lambda}\,|\,\phi_{\lambda^{\prime}}
\right > = \delta(\lambda - \lambda^{\prime}).
\end{equation}
With the aid of Eq.~(\ref{eq:n2}), it is easy to show that
\begin{equation}
\label{eq:n5}
\frac{\partial \lambda}{\partial t} = 0.
\end{equation}
The solutions $\psi_{\lambda} (q, \, t)$ for the Schr\"odinger
equation (\ref{eq:n1}) can be expressed as
\begin{equation}
\label{eq:n6} \psi_{\lambda} (q, \, t) = e^{i \alpha_{\lambda}(t)}
\phi_{\lambda}(q, \, t),
\end{equation}
where the phases functions $\alpha_{n}(t)$ are found from the equation
\begin{equation}
\label{eq:n7} \hbar \frac{\textrm{d}
\alpha_{\lambda}(t)}{\textrm{d} t}= \left <\phi_{\lambda} \left |
i \hbar \frac{\partial}{\partial t} - H(t) \right | \phi_{\lambda}
\right >.
\end{equation}
Therefore, we may write the general solution of Eq.~(\ref{eq:n1})
as[6]
\begin{equation}
\label{eq:n8}
\psi (q, \, t) = \int c(\lambda)\, e^{i
\alpha_{\lambda}(t)} \phi_{\lambda}(q, \, t) \, \textrm{d}\lambda,
\end{equation}
where
\begin{equation}
\label{eq:n9} c(\lambda) = \left <\psi_{\lambda}(q, \, 0) \, | \,
\psi(q, \, 0) \right >.
\end{equation}
Finally, let us recall that the calculations are similar, though
not identical to the discrete cases.
\\
\begin{center}
\textbf{III. THE GENERALIZED TIME-DEPENDENT INVERTED HARMONIC
OSCILLATOR}
\\
\textbf{A. Exact invariants and the solution of the Schr\"odinger
equation}
\end{center}
\par
Consider the generalized time-dependent inverted oscillator
described by the Hamiltonian
\begin{equation}
\label{eq:n10}
H(t) = \frac{p^2}{2 \, M(t)} -
\frac{1}{2}M(t)\omega^2(t)q^2 + \frac{y(t)}{2}(pq + qp),
\end{equation}
where $q$ e $p$ are canonical coordinates with $[q, \, p] = i
\hbar$, $M(t)$ and $\omega(t)$ are time-dependent mass and
frequency and $y(t)$ is an arbitrary function of time. Note that
the Hamiltonian (\ref{eq:n10}) can be obtained from the standard
generalized time-dependent oscillator by replacement $\omega(t)
\rightarrow i \omega (t)$. The Heisenberg's equations are
\begin{equation}
\label{eq:n11}
\dot{q} = \frac{1}{i \hbar} [q, \, H] =
\frac{p}{M(t)} + y(t) q,
\end{equation}
\begin{equation}
\label{eq:n12}
\dot{p} = \frac{1}{i \hbar} [p, \, H] =
M(t)\omega^2(t)q - y p.
\end{equation}
>From Eqs.~(\ref{eq:n11}) and (\ref{eq:n12}) we readily obtain that
\begin{equation}
\label{eq:n13}
\ddot{q} + \gamma(t) \dot{q} - \Omega^2(t)q = 0,
\end{equation}
where
\begin{equation}
\label{eq:n14}
\gamma(t) = \frac{\textrm{d}}{\textrm{d}t}
\textrm{ln}[M(t)]
\end{equation}
and
\begin{equation}
\label{eq:n15}
\Omega^2(t) = \omega^2 + y^2 + \gamma y + \dot{y},
\end{equation}
is the modified frequency. Now, it is easy to verify that an
invariant for the Hamiltonian (\ref{eq:n10}) is given by
\begin{equation}
\label{eq:n16} I(t) = \frac{1}{2} \left \{- \left ( \frac{q}{\rho}
\right )^2 + [\rho p - M (\dot{\rho} - y \rho)q]^2 \right \},
\end{equation}
where $\rho (t)$ is a $c$-number quantity satisfying the auxiliary equation
\begin{equation}
\label{eq:n17}
\ddot{\rho} + \gamma(t) \dot{\rho} -
\Omega^2(t)\rho = - \frac{1}{M^2 \rho^3}.
\end{equation}

Next, we want to solve the Schr\"odinger equation,
Eq.~(\ref{eq:n1}), with $H(t)$ given by [see Eq.~(\ref{eq:n10})]
\begin{equation}
\label{eq:n18}
H(t) = -\frac{\hbar^2}{2 \, M(t)}
\frac{\partial^2}{\partial q^2}- \frac{1}{2}M(t)\omega^2(t)q^2 - i
\hbar \frac{y(t)}{2} - i \hbar y(t) q \frac{\partial}{\partial q},
\end{equation}
where $p = - i \hbar \partial / \partial q$ has been used. To this
end, we consider the unitary transformation
\begin{equation}
\label{eq:n19}
\phi_{\lambda}^{\prime} (q, \, t) = U
\phi_{\lambda}(q, \, t),
\end{equation}
where
\begin{equation}
\label{eq:n20}
U = \exp \left [ - \frac{i M(t)}{2 \hbar \rho}
(\dot{\rho} - y \rho) q^2 \right ].
\end{equation}
Under this unitary transformation, the eigenvalue equation,
Eq.~(\ref{eq:n3}), with $I(t)$ given by Eq.~(\ref{eq:n16}) becomes
\begin{equation}
\label{eq:n21}
I^{\prime} \phi_{\lambda}^{\prime} (q, \, t) =
\lambda \phi_{\lambda}^{\prime} (q, \, t),
\end{equation}
with
\begin{equation}
\label{eq:n22}
I^{\prime} = U I U^{\dagger} = - \frac{\hbar}{2}
\rho^2 \frac{\partial^2}{\partial q^2} - \frac{1}{2} \left (
\frac{q}{\rho}\right )^2.
\end{equation}
Then, taking $\sigma = q/\rho$ we can write the eigenvalue
equation, Eq.~(\ref{eq:n21}), in the form
\begin{equation}
\label{eq:n23} \left [ -\frac{\hbar^2}{2}
\frac{\partial^2}{\partial \sigma^2} - \frac{\sigma^2}{2} \right ]
\varphi_\lambda (\sigma) = \lambda \varphi_{\lambda} (\sigma),
\end{equation}
or
\begin{equation}
\label{eq:n24} I^{\prime} \varphi_{\lambda}(\sigma) = \lambda
\varphi_{\lambda} (\sigma),
\end{equation}
where
\begin{equation}
\label{eq:n25} \phi_{\lambda}^{\prime} (q, \, t) =
\frac{1}{\rho^{1/2}} \varphi_{\lambda} (\sigma) =
\frac{1}{\rho^{1/2}} \varphi_{\lambda} (q/\rho).
\end{equation}
The factor $1/\rho^{1/2}$ is introduced into Eq.~(\ref{eq:n25}) so
that the condition
\begin{equation}
\label{eq:n26} \int {\phi_{\lambda}^{\prime}}^* (q, \, t)
\phi_{\lambda}^{\prime} (q, \, t) \, \textrm{d}q = \int
\varphi_{\lambda}^* (\sigma) \varphi_{\lambda} (\sigma) \,
\textrm{d}\sigma
\end{equation}
holds. Now, Eq.~(\ref{eq:n23}) is similar to the eigenvalue
equation of the time-independent inverted harmonic
oscillator\cite{r7}. Then, by setting
\begin{equation}
\label{eq:n27} \epsilon = \frac{\lambda}{\hbar}
\end{equation}
and
\begin{equation}
\label{eq:n28} z = \left ( \frac{2}{\hbar} \right )^{1/2}  \sigma,
\end{equation}
Eq.~(\ref{eq:n23}) can be written as
\begin{equation}
\label{eq:n29}
\frac{\partial^2 \varphi_{\lambda}}{\partial z^2} +
\left ( \frac{1}{4} z^2 + \epsilon \right ) \varphi_{\lambda} (z,
\, \epsilon) = 0.
\end{equation}
The solutions of Eq.~(\ref{eq:n29}) can be expressed in terms of
the parabolic cylinder (or Weber)
functions\cite{r3,r7,r8}\nocite{r8}. Thus, using
Eqs.~(\ref{eq:n19}), (\ref{eq:n20}), (\ref{eq:n25}),
(\ref{eq:n27}) and (\ref{eq:n28}) we get
\begin{equation}
\label{eq:n30}
\phi_{\lambda} (q, \, t) = \frac{1}{\rho^{1/2}}
\exp \left [ \frac{i M(t)}{2 \hbar \rho} (\dot{\rho} - y \rho) q^2
\right ] \varphi_{\lambda} \left [ \left ( \frac{2}{\hbar} \right
)^{1/2} \frac{q}{\rho}, \, \frac{\lambda}{\hbar} \right ],
\end{equation}
where $\varphi_{\lambda}$ in the Eq.~(\ref{eq:n30}) is the Weber's
function.

On the other hand, substituting Eqs.~(\ref{eq:n18}) and
(\ref{eq:n19}) into Eq.~(\ref{eq:n7}) and following the same steps
as those of Refs.\cite{r9,r10}\nocite{r9}\nocite{r10} we find that
the phase functions $\alpha_{\lambda} (t)$ are given by
\begin{equation}
\label{eq:n31} \alpha_{\lambda} (t) = -\frac{\lambda}{\hbar}
\int_0^t \frac{1}{M(t^{\prime}) \rho^2(t^{\prime})} \,
\textrm{d}t^{\prime} .
\end{equation}
Therefore, using Eqs.~(\ref{eq:n6}) and (\ref{eq:n30}) we obtain
that the exact solutions of the Schr\"odinger equation for the
generalized time-dependent inverted oscillator are
\begin{equation}
\label{eq:n32}
\psi_{\lambda} (q, \, t) = \frac{1}{\rho^{1/2}}
e^{i \alpha_{\lambda}(t)} \exp \left [ \frac{i M(t)}{2 \hbar \rho}
(\dot{\rho} - y \rho) q^2 \right ] \varphi_{\lambda} \left [ \left
( \frac{2}{\hbar} \right )^{1/2} \frac{q}{\rho}, \,
\frac{\lambda}{\hbar} \right ],
\end{equation}
where the phase functions are given by Eq.~(\ref{eq:n31}). Now,
the general solution of the Schr\"odinger equation is found by
directly substituting Eq.~(\ref{eq:n32}) into Eq.~(\ref{eq:n8}).
For $y(t) = 0$, the wave functions (\ref{eq:n32}) are exactly
reduced to those of the inverted oscillator with time-dependent
mass and frequency obtained in Ref.[4].
\begin{center}
\textbf{B. Generalized inverted oscillator with constant frequency and} \\
\textbf{exponentially increasing mass}
\end{center}
\par
In what follows, we consider the case where the frequency
$\omega(t)$ and the function $y(t)$ are constants, i.e.,
$\omega(t) = \omega_0$ and $y(t) = y_0$ and the mass is given by
\begin{equation}
\label{eq:n33} M(t) = m e^{\gamma t}.
\end{equation}
For this case, the Hamiltonian (\ref{eq:n10}) becomes
\begin{equation}
\label{eq:n34}
H(t) = e^{-\gamma t} \frac{p^2}{2 \, m} -
\frac{1}{2} m \omega_0 e^{\gamma t} q^2 + \frac{y_0}{2} (pq + qp),
\end{equation}
with $\gamma =$ const [see Eq.~(\ref{eq:n14})]. Note that in this
case the modified frequency $\Omega (t)$,  Eq.~(\ref{eq:n15}), is
constant and given by
\begin{equation}
\label{eq:n35}
\Omega^2(t) = \omega_0^2 + y_0^2 + \gamma y_0
\equiv \Omega_0^2.
\end{equation}
Here we observe that for $y_0 = 0$ the Hamiltonian (\ref{eq:n34})
reduces to that of the inverted Caldirola-Kanai
oscillator\cite{r3}.

Let us now consider a particular solution of Eq.~(\ref{eq:n17})
given by
\begin{equation}
\label{eq:n36} \rho (t) = \frac{1}{(m \Omega_1)^{1/2}} e^{-\gamma
t/2},
\end{equation}
where
\begin{equation}
\label{eq:n37}
\Omega_1^2 = \Omega_0^2 + \frac{\gamma^2}{4}.
\end{equation}
On the other hand, inserting Eqs.~(\ref{eq:n33}) and
(\ref{eq:n36}) into Eq.~(\ref{eq:n31}) we find that
\begin{equation}
\label{eq:n38}
\alpha_{\lambda} (t) = - \frac{\lambda
\Omega_1}{\hbar}t.
\end{equation}
Thus, by substituting Eqs.~(\ref{eq:n33}), (\ref{eq:n36}) and
(\ref{eq:n38}) into Eq.~(\ref{eq:n32}) we get after minor algebra
that
\begin{eqnarray}
\label{eq:n39} \psi_{\lambda} (q, \, t) = (m \Omega_1)^{1/4} \exp
\left [ \frac{\gamma t}{4} - i \frac{\lambda \Omega_1 t}{\hbar} -
\frac{i m e^{\gamma t}}{2 \hbar} \left ( \frac{\gamma}{2} + y_0
\right ) q^2 \right ] \nonumber \\ \times \varphi_{\lambda} \left
[ \left ( \frac{2 m \Omega_1}{\hbar} \right )^{1/2} e^{\gamma t/2}
q, \, \frac{\lambda}{\hbar} \right ],
\end{eqnarray}
which are the exact wave functions of the generalized inverted
oscillator with constant frequency and exponentially increasing
mass. For $y_0 = 0$, the wave functions (\ref{eq:n39}) are reduced
to those of the inverted Caldirola-Kanai oscillator\cite{r3,r4}.
$$\\
\textbf{IV. SUMMARY}
$$
\par
In this paper, we have discussed the LR invariant method for the
cases where the invariant has continuous eigenvalues and have
employed it to find the exact wave functions of a generalized
time-dependent inverted oscillator. We also have obtained, as a
special case, the wave functions of a generalized inverted
oscillator with constant frequency and exponentially increasing
mass. Furthermore, we have seen that for $y(t) = 0$ and $y_0 = 0$
our results agree with those obtained in Refs.\cite{r3,r4}.
Finally, we would like to remark that as far as we know the
Hamiltonians (\ref{eq:n10}) and (\ref{eq:n34}) and the wave
functions (\ref{eq:n32}) e (\ref{eq:n34}) have not yet been
exhibited in the literature.
$$\\
\textbf{ACKNOWLEDGMENTS}
$$
\par
Financial support from CNPq and FUNCAP Brazilian funding agencies
is gratefully acknowledged.

\end{document}